\documentclass[a4paper,10pt,twoside]{cpc-hepnp}
\usepackage{CJK,upgreek,fancyhdr}
\usepackage{multicol}
\usepackage{graphicx}
\usepackage{booktabs}
\usepackage{amssymb,bm,mathrsfs,bbm,amscd}
\usepackage[tbtags]{amsmath}
\usepackage{lastpage}
\usepackage{multirow}
\usepackage{color}
\usepackage{float}
\usepackage{lineno}

\begin{document}
\begin{CJK*}{GB}{gbsn}



\title{Studying the doublet bands at the borders of the $A \approx 130$ island of chiral candidates: the $^{120}$I case\thanks{Supported by National Natural Science
Foundation of China (11675063, 11205068, 11847310 and 11475072) }}

\author{Rui Guo~(¹ùÈð)$^{1}$
\quad Yong-Hao Liu~(ÁõÓÀºÃ)$^{1}$
\quad Jian Li~(Àî½£)$^{1,2;1)}$\email{jianli@jlu.edu.cn}
\quad Wu-Ji Sun~(ËïÎÞ¼É)$^{1}$
\quad Li Li~(˔ˏ)$^{1}$\\
\quad Ying-Jun Ma~(ÂíÓ¢¾ý)$^{1;2)}$\email{myj@jlu.edu.cn}}

\maketitle

\address{%
$^1$College of Physics, Jilin University, Changchun 130012, China\\
$^2$Department of Physics, Western Michigan University, Kalamazoo, MI 49008, USA}

\begin{abstract}
Based on the reported positive-parity doublet bands in $^{120}$I, the corresponding experimental characteristics including rotational alignment have been discussed and corresponding configuration assignment is reexamined.
  The self-consistent tilted axis cranking relativistic mean-field calculations indicates that this doublet bands is built on configuration $\pi h _{11/2}\otimes \nu h ^{-1}_{11/2}$.
  In addition, adopting the two quasiparticles coupled with a triaxial rotor model, the excitation energies, energy staggering parameter $S(I)$, $B(M1)/B(E2)$,
  effective angles, and $K$ $plots$ have been discussed and also compared with the available data.
  All of these results support the interpretation of chiral doublet bands for the positive-parity doublet bands in $^{120}$I, and hence identify this nucleus as the border of the $A \approx 130$ island of chiral candidates.
\end{abstract}



\footnotetext[0]{\hspace*{-3mm}\raisebox{0.3ex}{$\scriptstyle\copyright$}2019
Chinese Physical Society and the Institute of High Energy Physics
of the Chinese Academy of Sciences and the Institute
of Modern Physics of the Chinese Academy of Sciences and IOP Publishing Ltd}%

\begin{multicols}{2}
\section{Introduction}
Chirality is a topic of general interest in nature science, such as chemistry, biology, and physics. In nuclear structure physics, the occurrence of chirality was suggested for triaxially deformed
nuclei in 1997~\cite{Frauendorf1997Nucl.Phys.A131} and the predicted patterns of
spectra exhibiting chirality --- chiral doublet bands --- were
experimentally observed in 2001~\cite{Starosta2001Phys.Rev.Lett.971}, i.e., the existence of one pair of $\Delta I = 1$ nearly degenerate bands with the same parity.
Furthermore, the possible existence of two pairs or more of chiral doublet bands, i.e., multiple chiral doublet bands, in one nucleus was demonstrated and the acronym M$\chi$D was introduced by searching for triaxial chiral configurations in Rh isotopes based on constrained relativistic mean-field calculations in Ref.~\cite{Meng2006Phys.Rev.C037303}.
Up to now, nuclear chirality has become a hot topic in the current frontier of nuclear structure physics, and many chiral candidate nuclei, including several multiple doublet candidate nuclei~\cite{Peng2008PhysRevC.77.024309,Yao2009Phys.Rev.C067302,Li2011Phys.Rev.C037301,Qi2013PhysRevC.88.027302,Ayangeakaa2013PhysRevLett.110.172504,Kuti2014PhysRevLett.113.032501,Liu2016PhysRevLett.116.112501,ROY2018768,PhysRevC.98.014305,PhysRevC.97.034306,CHEN2018744,PhysRevC.98.024320}, have been reported
experimentally in the $A \sim 80, 100, 130$ and $190$ mass
regions of the nuclear chart; see e.g., the review Refs.~\cite{Meng2010J.Phys.G064025,Meng2016book} and data tables~\cite{XIONG2019193}.

\begin{center}
\includegraphics[width=8.5cm]{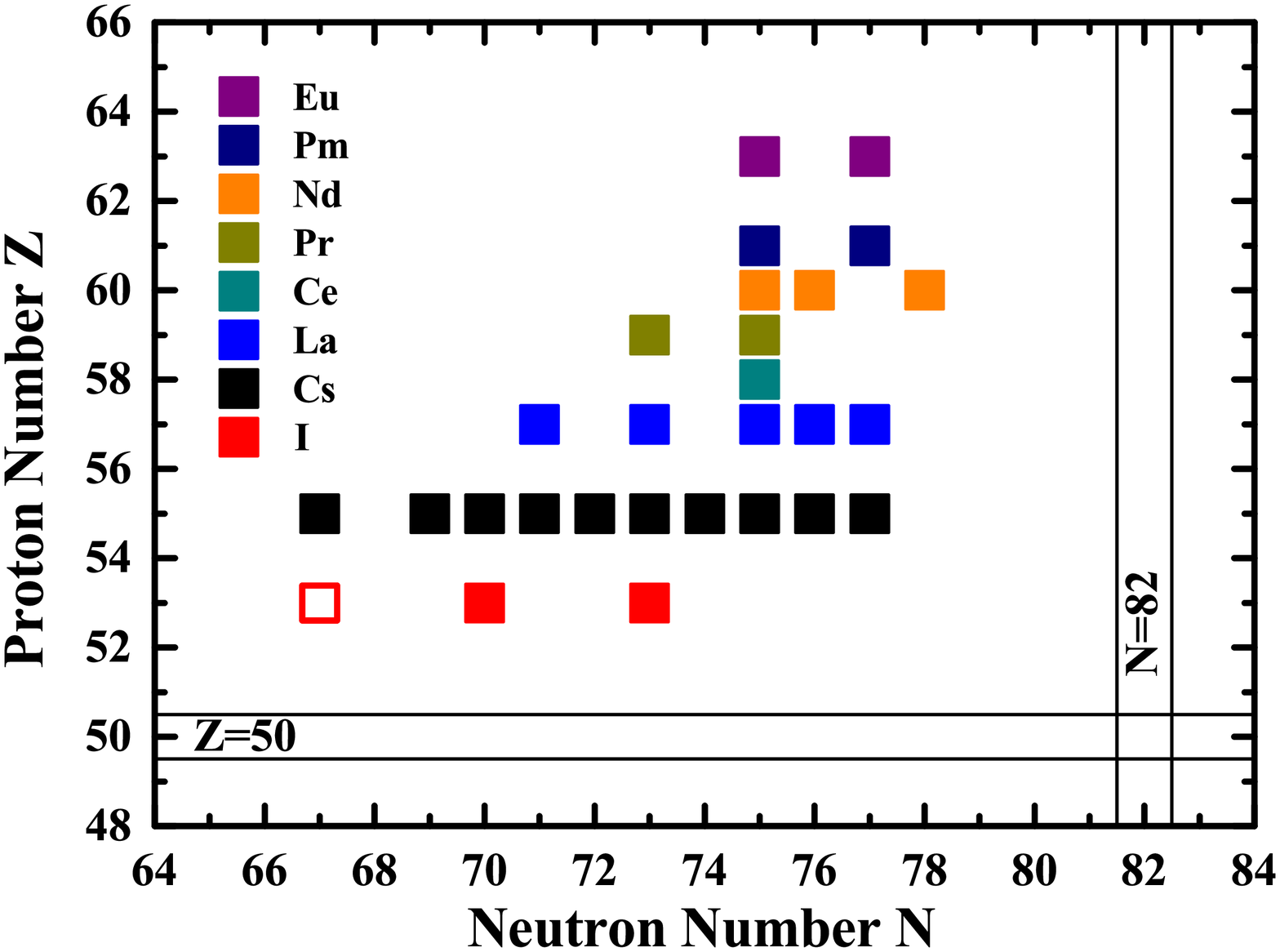}
\figcaption{
(Color online) \label{fig1} The reported candidate chiral nuclei in $A\sim130$ mass region, taken from
  Refs~\cite{Yong-Num2005J.Phys.GB1,Selvakumar2015PhysRevC.92.064307,Wang2006,Grodner2011Phys.Lett.B46,Grodner2006PhysRevLett.97.172501,Simons2005J.Phys.G541,Rainovski2003PhysRevC68.024318,Ma2012PhysRevC85.037301,Starosta2002PhysRevC65.044328,Bark2001Nucl.Phys.A577,Zhu2003PhysRevLett.91.132501,Merge2002Eur.Phys.J.A417,Petrache2012PhysRevC86.044321,Koike2001PhysRevC63.061304,Tonev2006,Hartley2001PhysRevC64.031304,Ayangeakaa2013PhysRevLett.110.172504,Hecht2001PhysRevC63.051302,Hecht2003PhysRevC68.054310,Yan_Xin_2009,Guo2019PhysRevC.100.034328,Zheng2012,Petrache2016,Ma2018}.}
\end{center}

Among the whole nuclear chart, most reported candidate chiral nuclei are located in $A\sim130$ mass region, where an island of chiral candidates was previously suggested in Ref.~\cite{Starosta2001Phys.Rev.Lett.971} and the available reported candidates in I, Cs, La, Ce, Pr, Nd, Pm and Eu~\cite{Yong-Num2005J.Phys.GB1,Selvakumar2015PhysRevC.92.064307,Wang2006,Grodner2011Phys.Lett.B46,Grodner2006PhysRevLett.97.172501,Simons2005J.Phys.G541,Rainovski2003PhysRevC68.024318,Ma2012PhysRevC85.037301,Starosta2002PhysRevC65.044328,Bark2001Nucl.Phys.A577,Zhu2003PhysRevLett.91.132501,Merge2002Eur.Phys.J.A417,Petrache2012PhysRevC86.044321,Koike2001PhysRevC63.061304,Tonev2006,Hartley2001PhysRevC64.031304,Ayangeakaa2013PhysRevLett.110.172504,Hecht2001PhysRevC63.051302,Hecht2003PhysRevC68.054310,Yan_Xin_2009,Guo2019PhysRevC.100.034328,Zheng2012,Petrache2016,Ma2018} could be seen in Fig.~\ref{fig1}.
Therefore, it is very interesting to investigate the boundary of the chiral island.
In $A\sim130$ mass region, the iodine isotopes lies on the lower edge of the present chiral island. In 2003, based on the similar energy spectrum, Moon et al.~\cite{MoonC-B2003} suggested that there may be chiral doublet bands in $^{120}$I.
In 2013, Li et al. extended both the yrast and side bands up to state $23^{+}$ and $17^{+}$ respectively~\cite{Li_2013}.
Recently, following the experimental properties of doublet bands in Ref.~\cite{MoonC-B2003}, Moon et al.~\cite{MOON2018602} discussed the chiral characteristics and doubling of states for the $\pi h_{11/2} \otimes \nu h_{11/2}$ configuration based on the  large-scale spherical shell model and total Routhian
energy surface calculations, and further suggested a chiral-like pair band in $^{120}$I.
Therefore, it is necessary to study the rotational mechanism for the occurrence of the chirality in $^{120}$I, and extend the boundary of the island of chirality.

On the theoretical side, different models of nuclear structure have been applied to study the nuclear chirality.
Among them, the particle rotor model (PRM)~\cite{Frauendorf1997Nucl.Phys.A131,Zhang2007PhysRevC.75.044307,Qi2009Phys.Lett.B175,Peng2003Phys.Rev.C044324,PhysRevLett.93.172502,PhysRevC.97.041303,PhysRevC.98.031303,PhysRevLett.120.022502,PhysRevC.99.064326,PENG2019303},
and the titled axis cranking (TAC) model including various versions~\cite{Frauendorf2001,Olbratowski2004Phys.Rev.Lett.052501,PhysRevC.73.054308,Zhao2017PhysLettB773,PhysRevLett.84.5732,PhysRevC.99.054319}, are most widely used.
In this article, the experimental characteristics and configuration assignments for the doublet bands in $^{120}$I, including energy differences, rotational alignment, and self-consistent tilted axis cranking relativistic mean-field (TAC-RMF) calculation~\cite{Peng2008Phys.Rev.C024313,Zhao2011Phys.Lett.B181,Meng2013,Zhao2017PhysLettB773}, are discussed first.
Then, the triaxial quasiparticles rotor model~(PRM)~\cite{Zhang2007PhysRevC.75.044307} will be adopted to investigate the rotational structure and possible chirality in the doublet bands. Finally, a brief summary is given.
\end{multicols}

\begin{multicols}{2}
\section{Experimental characteristics and configuration
assignments for the doublet bands}
\begin{center}
\includegraphics[width=9.5cm]{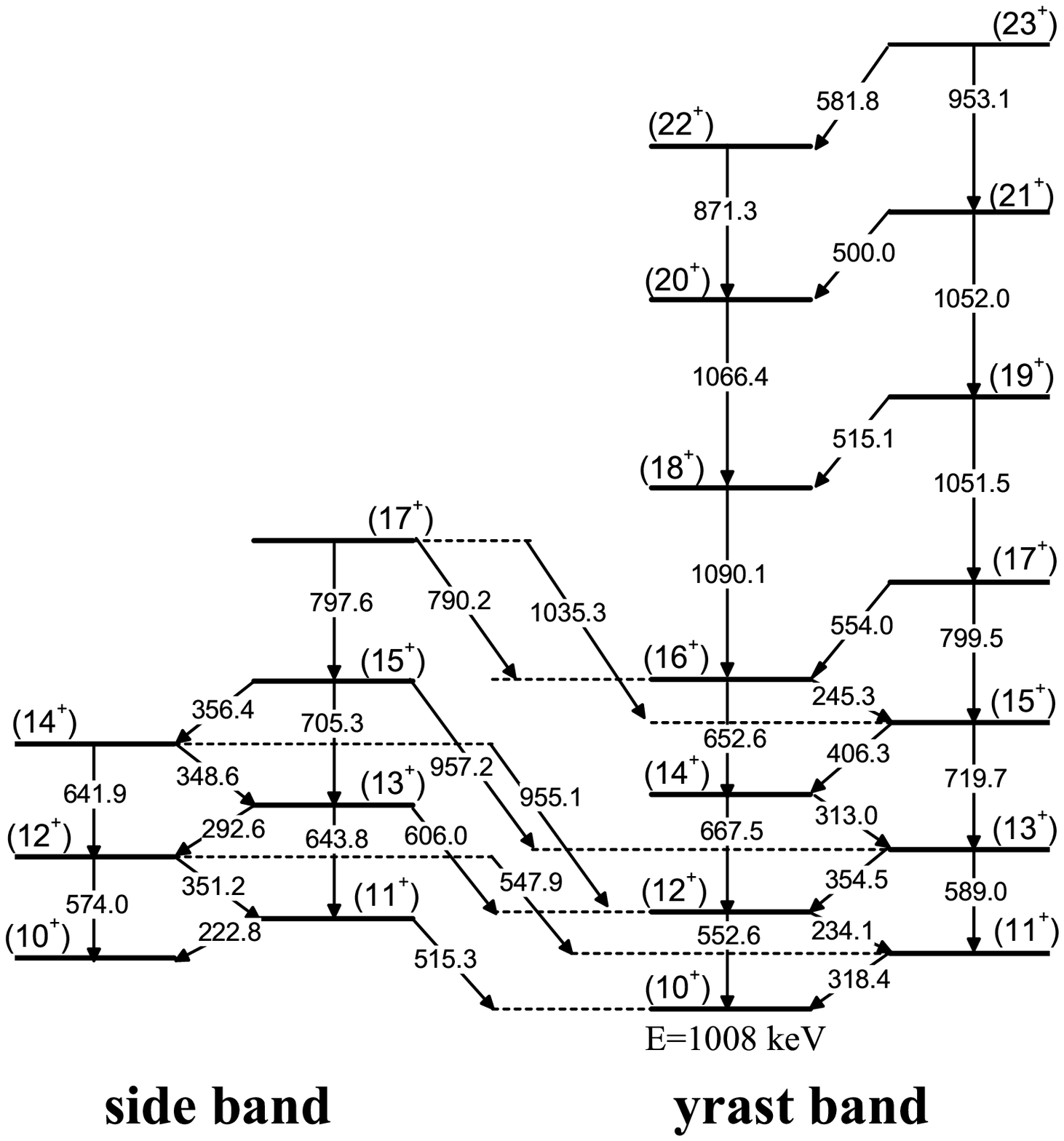}
\figcaption{\label{fig2} Partial level scheme of doublet bands in $^{120}$I including the linking transitions. Full level schemes could be found in Refs.~\cite{Li_2013,L.LiPHDthesis}. All the energy scales are in keV.}
\end{center}

Partial level scheme for $^{120}$I derived from Refs.~\cite{Li_2013,L.LiPHDthesis} is shown in Fig.~\ref{fig2}, where the
yrast band and the side band, as well as the linking transitions between them are adopted from Ref.~\cite{Li_2013,L.LiPHDthesis}.
The experiment was performed at the HI-13 tandem accelerator of the China Institute of Atomic Energy, and high-spin states in $^{120}$I were populated by using the fusion-evaporation reaction $^{114}$Cd($^{11}$B,5n) at
a beam energy of 70 MeV. The yrast band built on the ($10^+$) level at 1008 keV has been extended up to the $(23^+)$ state, while the side band also built on the ($10^+$) state and is extended up to ($17^+$).
Apart from the similarity of energy spectra for the doublet bands, the existence of linking transitions multipolarities $E2$ and $M1/E2$ between the two bands also support the judgment that the side band has the same parity and configuration as that of the yrast band.

To study the characteristics of the doublet bands, the rotational alignment of the doublet bands in $^{120}$I are shown in Fig.~\ref{fig3}, compared with the neighboring chiral doublet bands in $^{122}$Cs~\cite{Yong-Num2005J.Phys.GB1} and $^{124}$Cs~\cite{Selvakumar2015PhysRevC.92.064307}. To subtract the angular momentum of the core, the Harris parameters~\cite{PhysRevC.44.2390}, i.e., $J_{0}=17.0\hbar^{2}/$MeV,
$J_{1}=25.8\hbar^{4}/$MeV$^{3}$, are adopted.
In Fig.~\ref{fig3}, the large initial alignments ($\sim$$6\hbar$) for yrast band and side band in $^{120}$I indicate that the $h_{11/2}[550]1/2^{-}$ proton
which can donate large alignment ($\sim$$5.5\hbar$) is involved in the configuration of yrast band and side band.
In the neighboring isotone $^{121}$Xe, the $h_{11/2}$ neutron rotational band is closer to the yrast line than the $d_{5/2}$ and $d_{3/2}$ neutron rotational band~\cite{Timar_1995}.
Therefore, the odd neutron in $^{120}$I may have the priority to occupy
the $h_{11/2}$ orbit, and the same configuration $\pi h_{11/2}\otimes \nu h_{11/2}$ as the assignments in Ref.~\cite{MOON2018602} is favored for the doublet bands in $^{120}$I.

On the other side, it can be seen from Fig.~\ref{fig3} that the alignment values of the doublet bands in $^{120}$I
are very close to the alignment values of the doublet bands in $^{122}$Cs and $^{124}$Cs, during the rotational frequency between 0.25 and 0.34 MeV.
This indicates that the configurations of the doublet bands in $^{120}$I in the frequency range from 0.25 to 0.34 MeV should be same as that of doublet bands in $^{122}$Cs and $^{124}$Cs, i.e., the high-$j$ $h_{11/2}$ proton particle and neutron hole configuration.
Therefore, these doublet bands have similar band structure and angular momentum alignment, which support that they have the same configuration.

\begin{center}
\includegraphics[width=8.5cm]{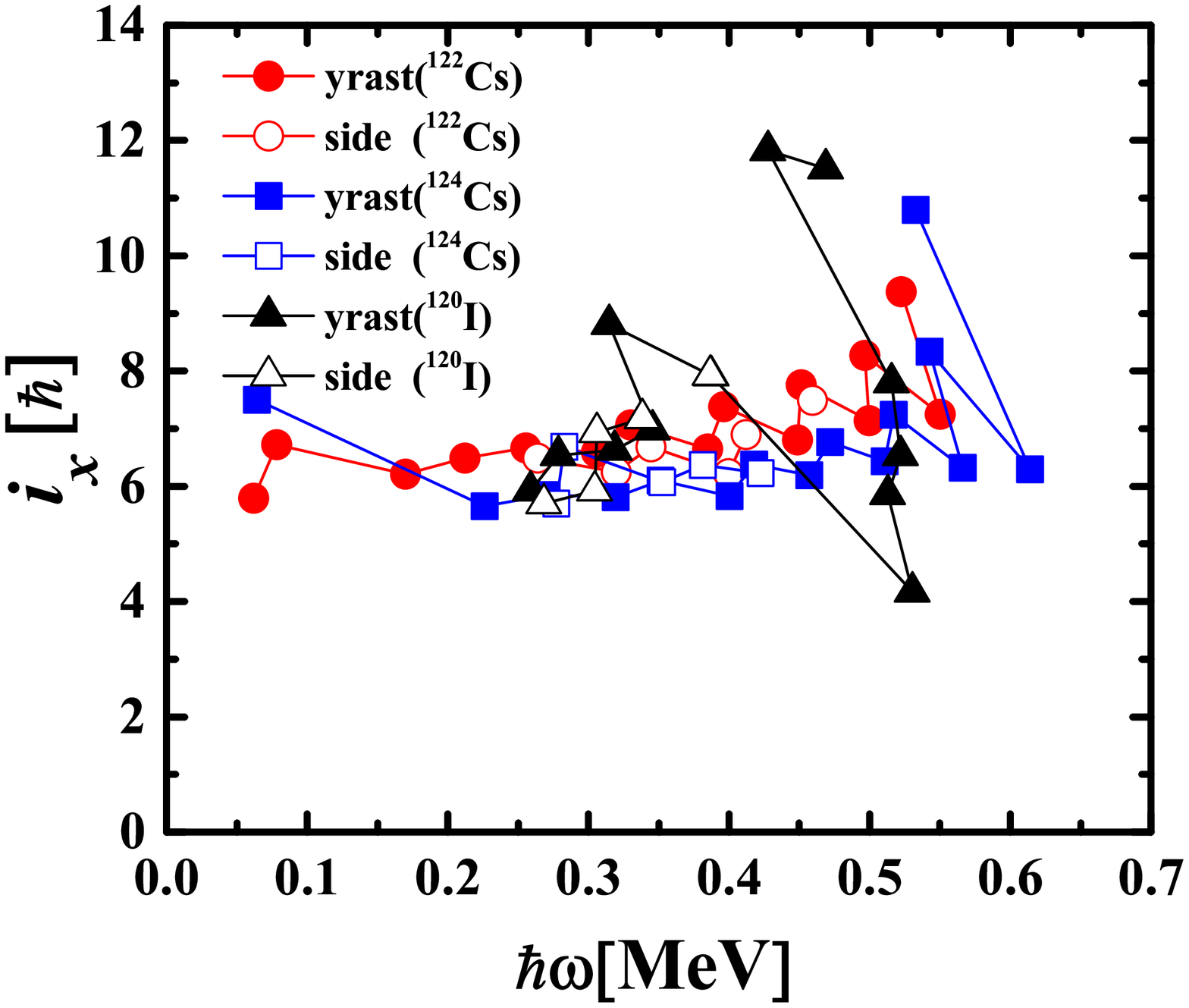}
\figcaption{
(Color online) \label{fig3} Rotational alignment of yrast and side bands in $^{120}$I, compared with $^{122}$Cs and $^{124}$Cs. Harris parameters~\cite{PhysRevC.44.2390}($J_{0}=17.0\hbar^{2}/$MeV,
  $J_{1}=25.8\hbar^{4}/$MeV$^{3}$) are used to subtract the angular momentum of the core.}
\end{center}

In addition, it can be noticed that the alignment of yrast band in $^{120}$I suddenly increases by about 2 $\hbar$ at around the rotational frequency 0.34 MeV,
and then gradually decreases until 0.53 MeV where the backbending phenomenon occurs.
It may be owing to non-aligned proton $\pi g_{7/2}$ and neutron $\nu h_{11/2}$ excitation in the core.
Such excitation has been observed in the $^{118}$Te, i.e., the core of $^{120}$I, and its isotope $^{120}$Te~\cite{118TePhysRevC.61.014312,PhysRevC.85.014310}.
When the paired $g_{7/2}$ protons are excited, the deformation of nucleus for $^{120}$I begins to change and gradually becomes oblate, and the moment of inertia becomes smaller,
which is likely the reason for the decrease of alignment value at the rotational frequencies about 0.3-0.5 MeV, as discussed in Ref.~\cite{L.LiPHDthesis}.

To further investigate the configuration of the doublet bands in $^{120}$I, the TAC-RMF calculations have been performed.
During the past few decades, the relativistic mean field (RMF) theory has been a great success in describing properties of nuclei and many nuclear phenomena~\cite{Ring1996Prog.Part.Nucl.Phys.193,Vretenar2005Phys.Rep.101,Meng2006Prog.Part.Nucl.Phys.470,Meng2016book,RevModPhys.75.121}. Based on the RMF theory, the tilted axis cranking relativistic
mean-field (TAC-RMF) theory has been developed for
describing many nuclear rotational phenomenon such as magnetic, antimagnetic and chiral rotation~\cite{Meng2013,Meng2016book,Zhao2017PhysLettB773,Sun2019Chin.Phys.C,Guo2019Phys.Rev.C}.
In the TAC-RMF theory, nuclei are characterized by the relativistic fields $S(\bm{r})$ and $V^{\mu}(\bm{r})$ in the Dirac equation in the rotating frame with a constant angular velocity vector $\bm{\Omega}$ as
\begin{equation}\label{DiracEq}
 [\bm{\alpha}\cdot(-i\nabla- \bm{V} ) +\beta (m + S)+V - \bm{\Omega}\cdot \bm{\hat{J}}]\psi_i =\varepsilon_i\psi_i,
\end{equation}
where $\bm{\hat{J}}=\bm{\hat{L}}+\frac{1}{2}\bm{\hat{\Sigma}}$ is the total angular momentum of the nucleon spinors, and $\varepsilon_i$ represents the single-particle
Routhians for nucleons. The detailed formalism and numerical techniques can be seen in Refs.~\cite{Peng2008Phys.Rev.C024313,Zhao2011Phys.Lett.B181,Zhao2012Phys.Rev.C054310}. A spherical harmonic oscillator basis with ten major shells is adopted to solve the Dirac equation. The point-coupling interaction PC-PK1~\cite{Zhao2010Phys.Rev.C054319} is used for the Lagrangian. Due to the suppression of pairing effects for the
high $j$ two-quasiparticle configuration, the pairing correlations are neglected for simplicity here.
It should be noted that the pairing correlation implemented in a fully self-consistent and microscopic way in TAC-RMF plays
an important role on the description of energy spectrum and electromagnetic transitions in nuclear rotation~\cite{Zhao2015PhysRevC.92.034319,PhysRevC.97.064321,PhysRevC.97.034317,PhysRevC.99.024317}.
Therefore, it is very desiring the unified and self-consistent investigations of nuclear chirality within the tilted axis cranking
covariant density functional theory with the pairing correlation in the future.

\begin{center}
\includegraphics[width=8.5cm]{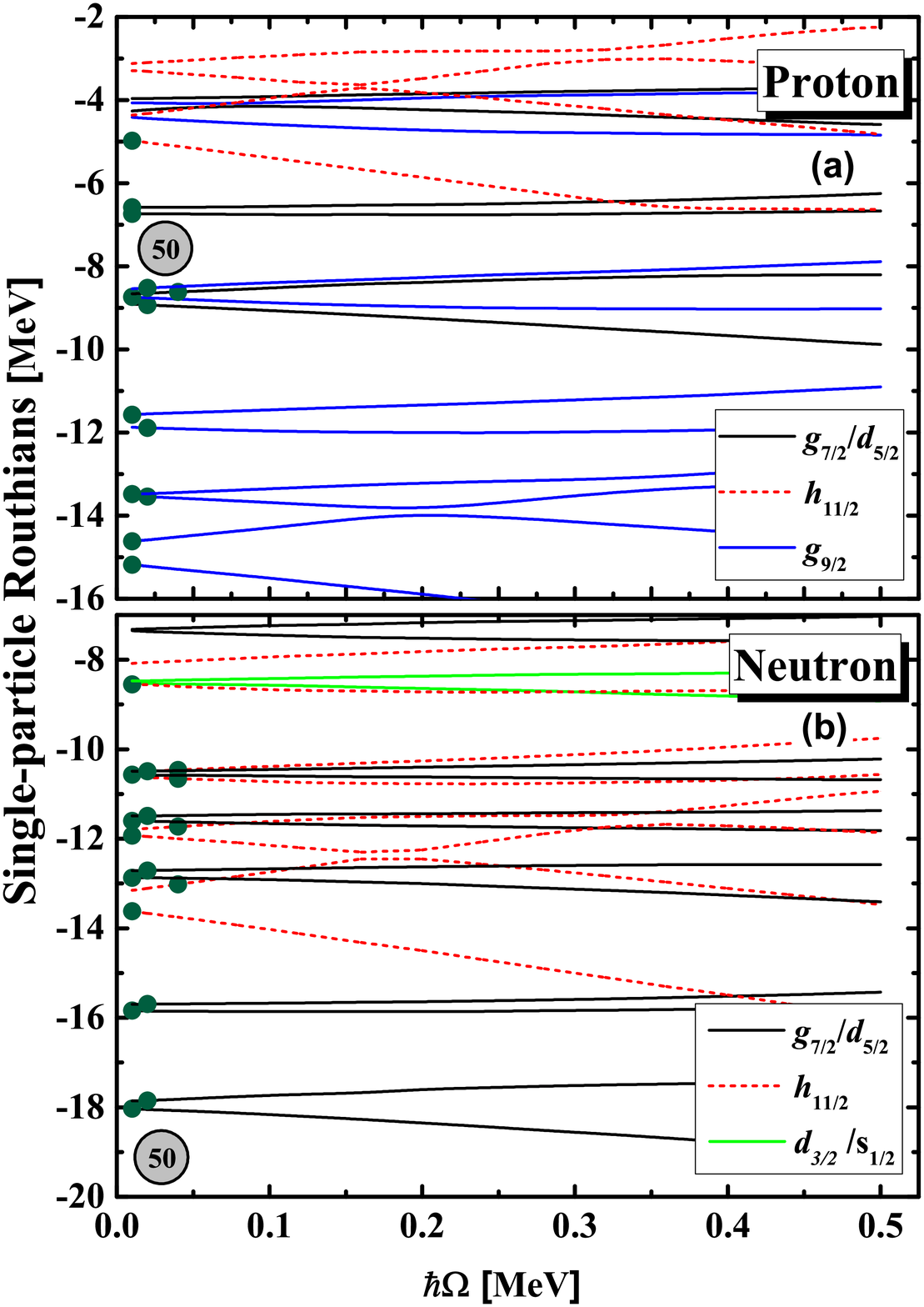}
\figcaption{
(Color online) \label{fig4} Single-proton (upper panel) and single-neutron (lower panel) Routhians near the Fermi surface in $^{120}$I as a function of the rotational frequency for configuration $\pi h_{11/2}\otimes\nu h^{-1}_{11/2}$ in TAC-RMF calculation.
  The levels of $\nu{g_{7/2}}/{d_{5/2}}$, $\nu h_{11/2}$, $\nu{d_{3/2}/s_{1/2}}$ and $\pi{g_{9/2}}$ are marked by solid black, dashed red, solid green and blue lines, respectively.
  The filled olive circles indicate the occupied levels.}
\end{center}

In Fig.~\ref{fig4}, the single-particle Routhians for the protons (a) and neutrons (b) in $^{120}$I as a function of rotational frequency are shown. In principle, broken time-reversal symmetry by the cranking field will be recovered without cranking field at $\Omega = 0$, i.e., the levels will be degenerated again as already shown in Fig.~\ref{fig4}.
There are three protons and seventeen neutrons above the $N = 50$ shell in $^{120}$I.
From Fig.~\ref{fig4}(a), it is easy to see that the last unpaired proton is occupying in the lower part of $h_{11/2}$ in ascending order of energy.
From Fig.~\ref{fig4}(b), it can be seen that the last unpaired neutron is kept fixed in the upper part of $h_{11/2}$ for a large range of rotational frequency.
The other remaining sixteen neutrons are treated self-consistently by filling the orbits according to their energies and distributed over the $h_{11/2}$ and ${g_{7/2}}/{d_{5/2}}$ shells, respectively.
In short, the high $j$ proton-particle and neutron-hole configurations favored the doublet bands in $^{120}$I,
whereas seven neutrons occupying the $h_{11/2}$ orbital means the character of neutron-hole is not ideal,
which may be the reason that the valence neutron can not be obviously considered as a hole in character as discussed in Ref.~\cite{MOON2018602}.

It can be deduced from Fig.~\ref{fig4} that when the neutron number decreases in the iodine isotopes, i.e., the neutron number $N$ less than 77, the last odd neutron will occupy in the lower part of $h_{11/2}$ shell and neutron $h_{11/2}$ hole character may not be favored. Similarly, when the proton number $Z$ is less than 53, there will be no proton occupying in the $h_{11/2}$ orbit, and high-$j$ $h_{11/2}$ proton particle also can not be expected. Considering the proper high-$j$ particle and hole configuration as a necessary condition for the appearance of the chiral doublet bands, $^{120}$I could be the edge of island of chiral candidates in $A\sim130$ mass region. At the same time, it is naturally interesting to find and predict the nuclear chirality in more neighbouring nuclei of $^{120}$I.

\end{multicols}

\begin{multicols}{2}

\section{Chiral structure in particle rotor model}
\begin{center}
\includegraphics[width=8.5cm]{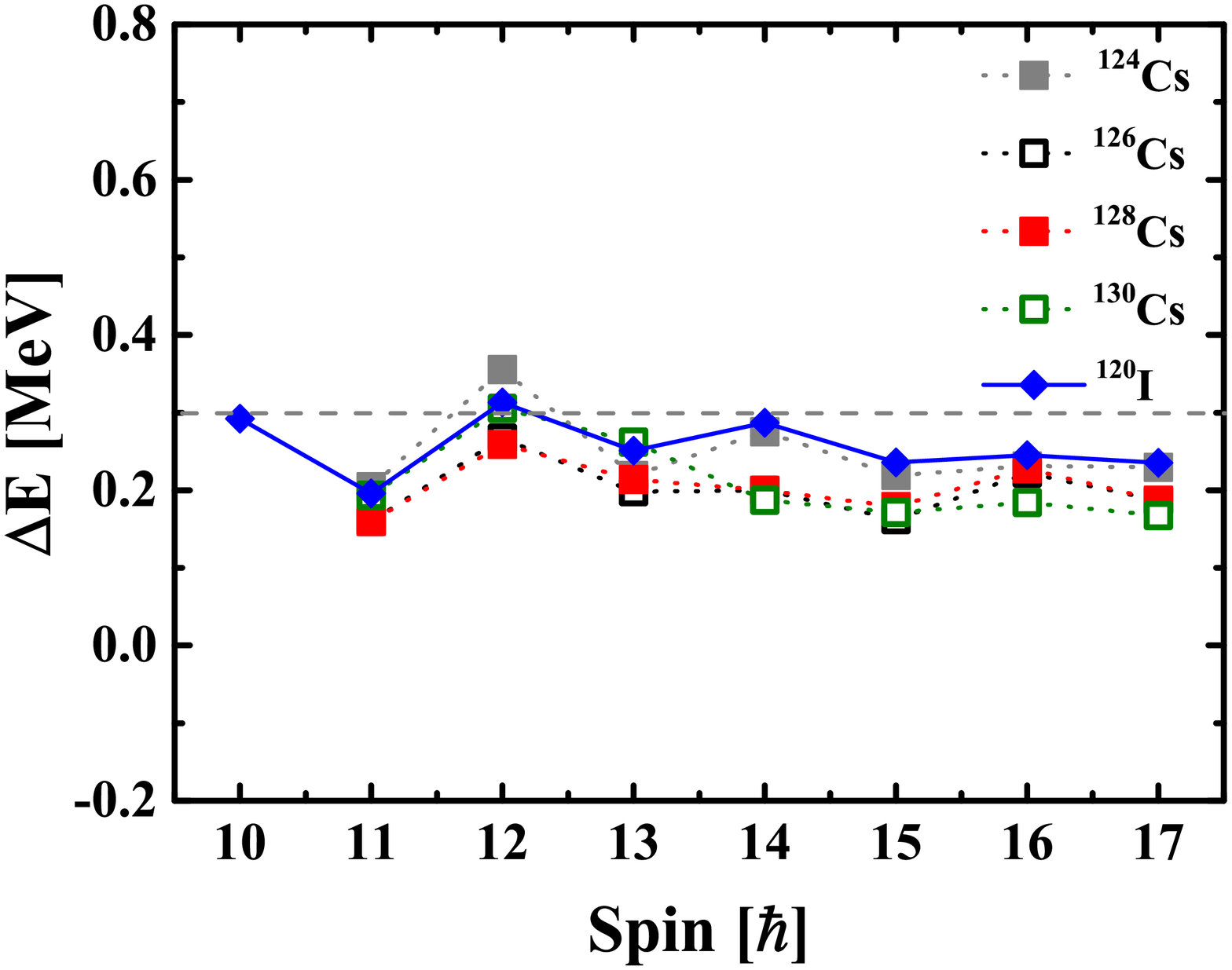}
\figcaption{
(Color online) \label{fig5} Comparison of energy differences $\triangle{E}$, i.e., $E(I)_{side}-E(I)_{yrast}$ as a function of the spin for the doublet bands in $^{120}$I and $^{124,126,128,130}$Cs.}
\end{center}

The present configuration assignments of high $j$ proton-particle and neutron-hole configurations for the doublet bands in $^{120}$I are favorable for the chirality.
It can be found that configuration components are almost the same components of chiral bands in the neighboring odd-odd nuclei $^{124-128}$Cs.
For this reason, a systematic comparison of their experimental features is performed.
In Fig.~\ref{fig5}, the energy differences $\Delta{E}$ for the doublet bands,
i.e., $E(I)_{side}-E(I)_{yrast}$, in $^{120}$I and $^{124-128}$Cs are shown.
It can be seen that the energy differences for these doublet bands are very similar.
They exhibit a slight undulation and gradually decrease from 360 to 160 keV with the spin increases from 14 to $17\hbar$.
$^{124}$Cs is especially similar to $^{120}$I, i.e., the energy differences $\Delta{E}$ is almost the same (greater than 200 keV) and is relatively large than to $^{126,128,130}$Cs, which implies $^{120}$I and $^{124}$Cs~\cite{PhysRevC.82.027303} may have a similar chiral geometry,
i.e., might correspond to a typical chiral vibration pattern.
Furthermore, the gradual decrease in energy differences of $^{120}$I may indicate that the chiral vibration patterns of the nucleus gradually form and tend to stabilize the chirality.

To investigate the existence of nuclear chirality in $^{120}$I, the particle rotor model (PRM) with a quasi-proton and a quasi-neutron coupled with a triaxial rotor~\cite{Zhang2007PhysRevC.75.044307} has been adopted. PRM, as a quantal model describing a system in the laboratory reference frame and consisting of collective rotation as well as intrinsic single particle motions,
can be straightforwardly applied to investigate the angular momentum geometries of chiral doublet bands.
In PRM, the total Hamiltonian is diagonalized with total angular momentum as a good quantum number, and the energy splitting and quantum tunneling between the doublet bands can be obtained directly.
Its Hamiltonian for an odd-odd nucleus can be written as~\cite{Zhang2007PhysRevC.75.044307}
\begin{eqnarray}
\label{eq2}
H= H_{coll}+H^{p}_{intr}+H^{n}_{intr},
\end{eqnarray}
where $p$ and $n$ refer to proton and neuron, respectively.
In addition,the configuration of multi-particles sitting in
a high-$j$ shell can be simulated by adjusting the Fermi energy here with considering pairing correlation by BCS approximation~\cite{Zhang2007PhysRevC.75.044307}. For detailed formalism of PRM see Refs.~\cite{Frauendorf1997Nucl.Phys.A131,Zhang2007PhysRevC.75.044307,Qi2009Phys.Lett.B175,Peng2003Phys.Rev.C044324,PhysRevLett.93.172502}.

The quadrupole deformation parameter $\beta=0.36$ for the configuration $\pi h _{11/2}\otimes \nu h ^{-1}_{11/2}$ have been adopted, which is obtained from both triaxial relativistic mean-field~(RMF)~\cite{Meng2006Phys.Rev.C037303,Peng2008PhysRevC.77.024309,Li2011Phys.Rev.C037301} and TAC-RMF calculations~\cite{Peng2008Phys.Rev.C024313,Zhao2011Phys.Lett.B181,Sun2016ChinPhysC.40.084101}.
At the same time, the triaxial deformation parameter $\gamma=20^{\circ}$ is employed to achieve better description of the experimental ratios of reduced transition probabilities $B(M1)/B(E2)$.
The single-$j$ shell Hamiltonian parameter was taken as~\cite{Shou_Yu_2009}
\begin{eqnarray}
\label{eq3}
C= \Big(\frac{123}{8}\sqrt{\frac{5}{\pi}}\Big) \frac{2N+3}{j(j+1)}A^{-1/3}\beta.
\end{eqnarray}

In addition, the moment of inertia $\mathcal{J}=25\hbar^{2}/$MeV
is adjusted according to the experimental energy spectra. Following the empirical formula $\Delta=12/\sqrt{A}$, the pairing gap $\Delta=1.1$ MeV is used for both protons and neutrons. As the valence proton Fermi level in $^{120}$I is supposed to be at the
beginning of $\pi h_{11/2}$ subshell, the proton Fermi energy $\lambda_{p}$ takes $-3.518$ MeV. Considering the neutron Fermi level lying in the middle of the $\nu_{11/2}$ subshell, the neutron Fermi energy $\lambda_{n}$ takes 0.765 MeV with simulating the effect of multivalence neutrons, similar as in the Refs.~\cite{Zhang2007PhysRevC.75.044307,PhysRevC.75.024309}. For the electromagnetic transition,
the empirical intrinsic quadrupole moment $Q_{0}=(3/\sqrt{5\pi})R_{0}^{2}Ze\beta=4.3$ eb. Gyromagnetic ratios for the collective rotator is given by $g_{R}=Z/A=0.44$. $g_{p}=1.21$ and $g_{n}=-0.21$ are adopted for proton and neutron respectively obtained from the Schmidt magnetic moment formula of $h_{11/2}$ orbital with effective spin gyromagnetic ratio $g_s=0.6\,g_s^{\mathrm{free}}$ used same as the Refs.~\cite{Qi2011,CHEN2018744,PENG2019303}.

\begin{center}
\includegraphics[width=8.5cm]{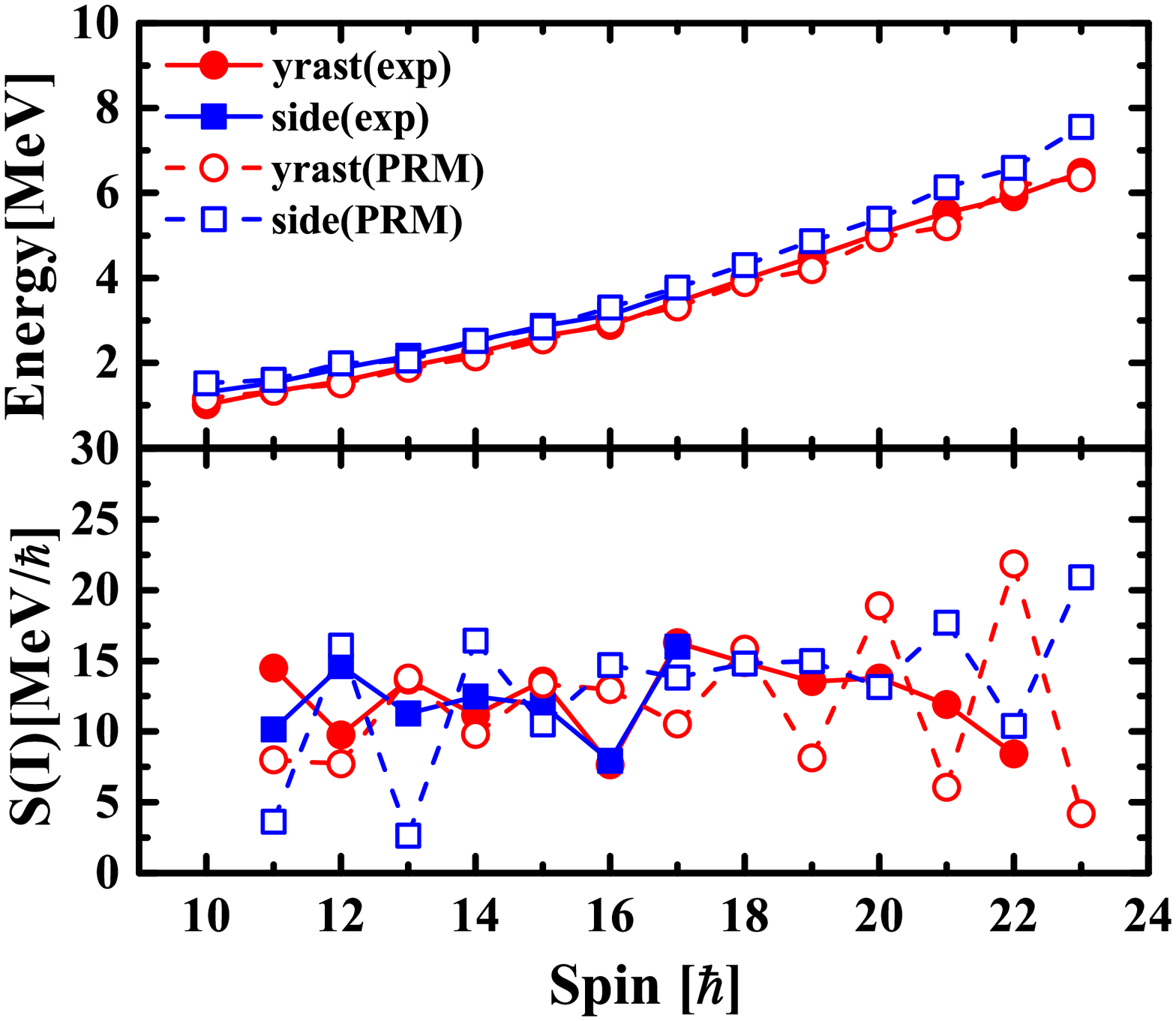}
\figcaption{
(Color online) \label{fig6} The excitation energy (upper panel) and staggering parameter $S(I)=[E(I)-E(I-1)]/2I$ (lower panel) as a function of spin for the yrast and side band in $^{120}$I.
  The filled (open) symbols connected by solid (dashed) lines indicate the experimental (theoretical) values. The yrast and side band are shown by circles and squares, respectively.}
\end{center}

In Fig.~\ref{fig6}, the calculated energy spectra $E(I)$ and energy staggering parameter $S(I)$, i.e., $[E(I)-E(I-1)]/2I$, within the PRM calculations for the doublet bands in $^{120}$I are presented,
in comparison with the corresponding experimental results.
One can easily see that the calculated energy spectra and energy staggering parameter $S(I)$ well reproduce the experimental results in the spin region $10-15\hbar$.
One of experimental fingerprint for nuclear chirality is that $S(I)$ varies smoothly with the increasing spin, which results from a highly reduced Coriolis interaction~\cite{Joshi2007PhysRevLett.98.102501}.
As shown in Fig.~\ref{fig6}, $S(I)$ exhibits a fairly smooth variation, supporting the nuclear chirality for the present doublet bands~\cite{MOON2018602}.
In addition, it can be noted from Fig.~\ref{fig6} that $S(I)$ of the doublet bands gradually approaches to each other with the increasing spin, and exhibits a slight staggering which gradually becomes weak.
Moreover, the behaviors of $S(I)$ are similar to that for the chiral bands in $^{124,126,128,130}$Cs,
indicating that the Coriolis interaction may reduce little by little, thereby suggesting an aplanar rotation gradually appears in the triaxial nuclei.
In other words, the chiral vibration pattern of the nucleus may gradually form and approach to static chirality.

\begin{center}
\includegraphics[width=8.5cm]{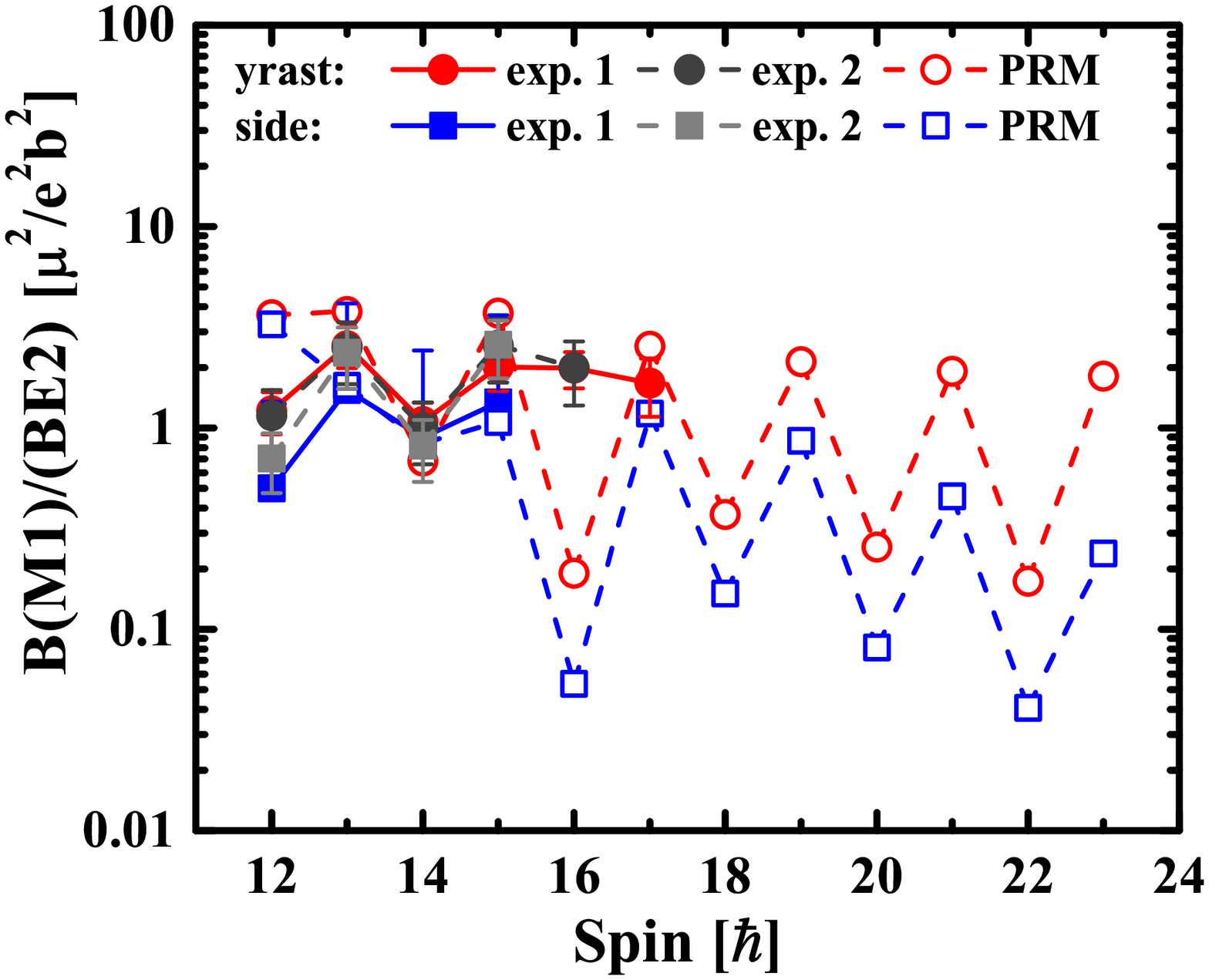}
\figcaption{
(Color online) \label{fig7} The calculated $B(M1)/B(E2)$ and the corresponding available data for the yrast and side band in $^{120}$I. The data extracted from Ref.~\cite{L.LiPHDthesis} is marked as exp. 1 and the corresponding data from Ref.~\cite{MOON2018602} is marked as exp. 2.}
\end{center}

The calculated in-band $B(M1)/B(E2)$ ratios and the corresponding available data extracted from Ref.~\cite{L.LiPHDthesis} (marked as exp. 1) for the yrast and side band in $^{120}$I are presented in Fig.~\ref{fig7}.
The corresponding data from ref.~\cite{MOON2018602} (marked as exp. 2) is also given for comparison.
It can be seen that the $B(M1)/B(E2)$ ratios of the doublet bands from Ref.~\cite{L.LiPHDthesis} and Ref.~\cite{MOON2018602} are very close to each other,
and both meet the important experimental criteria of chiral double bands, that is, the $B(M1)/B(E2)$ ratios of chiral doublet bands should be similar and show staggering behavior with increasing spin.
Besides, the feature of the $B(M1)/B(E2)$ is reproduced well by the present PRM calculation including the magnitude and staggering of the ratios with spin.
The success in reproducing the energy spectra,
the energy staggering parameter $S(I)$ and transition probabilities for the doublet bands in $^{120}$I suggests that the present calculation must correctly account for the
structure of the states at the spin region $10-15\hbar$.

To shed light on investigating the chiral picture of $^{120}$I, the effective angels $\theta_{pn}$, $\theta_{Rp}$, and $\theta_{Rn}$ obtained from the PRM calculations as a function of the spin for the doublet bands are displayed in Fig.~\ref{fig8}.
In the PRM, the effective angle $\theta$ refers to the angle between the angular momentum orientations of the nuclei in the intrinsic reference frame.
For instance, the effective angle $\theta_{pn}$, which refers to the angle between the proton (${\mathbf{j}_{p}}$) and neutron (${\mathbf{j}_{n}}$) angular momenta, is defined as~\cite{Starosta2002PhysRevC65.044328}
$\cos\theta_{pn}=\langle\mathbf{j}_{p}\cdot\mathbf{j}_{n}\rangle/\sqrt{(j_{p}^{2})(j_{n}^{2})}$,
and similarly for $\theta_{Rp}$ and $\theta_{Rn}$. Here, the subscripts $p, n$, and $R$ denote the proton, neutron, and rotor, respectively.
By analyzing the effective angles, the chiral geometry represented by a remarkable and similar aplanar rotation between doublet bands can be revealed in a quantum way.

\begin{center}
\includegraphics[width=8.5cm]{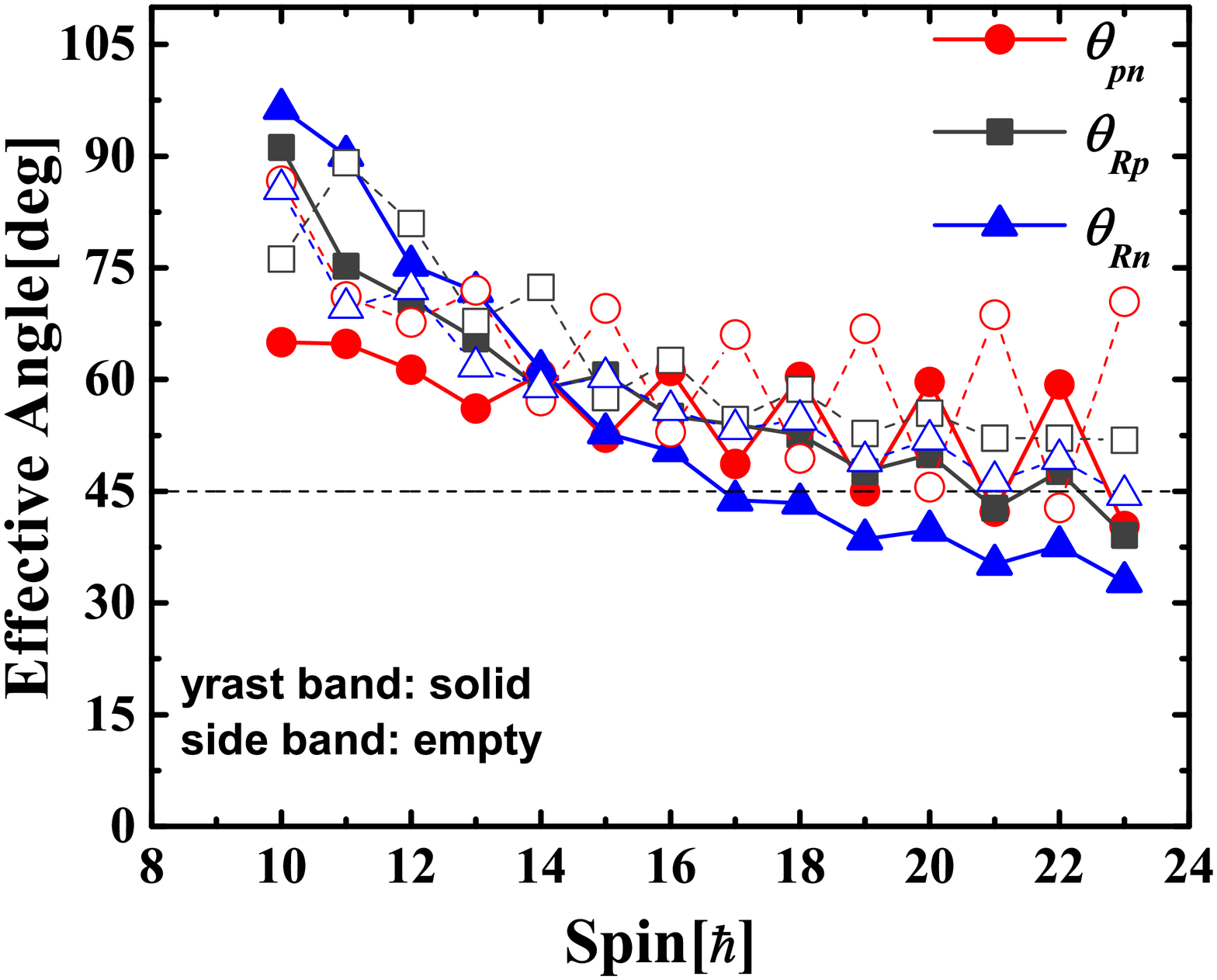}
\figcaption{
(Color online) \label{fig8} The effective angles $\theta_{pn}$, $\theta_{Rp}$, and $\theta_{Rn}$ as a function of the spin for the doublet bands in $^{120}$I. See text for more details.}
\end{center}

From Fig.~\ref{fig8}, it is observed that the effective angles $\theta_{pn}$, $\theta_{Rp}$, and $\theta_{Rn}$ are greater than $60^{\circ}$ around $I=10\hbar$, i.e., the angular momenta ${\textbf{j}_{p}}$,
${\textbf{j}_{n}}$, and $\mathbf{R}$ are nearly perpendicular to each other at the band head, which also appears in $^{124,126}$Cs~\cite{PhysRevC.82.027303,PhysRevC.75.024309}.
In the spin region $10-15\hbar$, the values of three effective angles for both yrast and side bands are larger than $45^{{\circ}}$,
which indicates a obvious aplanar rotation in $^{120}$I and provides additional support for the existence of chiral doublet bands in $^{120}$I.

For the sake of exhibiting the chiral picture more clearly and examine the evolution of the chiral geometry with angular momentum,
the probability distributions for the projection $\emph{K}$ of the total angular momenta ($\emph{K plots}$) of the doublet bands in $^{120}$I are calculated similarly as in Refs.~\cite{Qi2009Phys.Lett.B175,Qi2009PhysRevC.79.041302,PhysRevC.82.027303,PhysRevC.100.054309,Qi2009Phys.Lett.B175}.
The $\emph{K plots}$, i.e, the probability distributions for the projection $\emph{K}$ of total angular momenta along long ($\emph{l}$), intermediate ($\emph{i}$), and short ($\emph{s}$) axes are displayed in Fig.~\ref{fig9}.
At the band head $I=10\hbar$, the probability distributions along the $\emph{i}$ axis, i.e., $K_{\emph{i}}$ for the doublet bands are different.
For yrast band the maximum probability $K_{\emph{i}}$ appears at $K_{\emph{i}}=0$, whereas the $K_{\emph{i}}$ for side band is minimum at $K_{\emph{i}}=0$, having its peak at $K_{\emph{i}}\approx8$.
This is in accordance with the interpretation of the chiral vibration with respect to the $\emph{s-l}$ plane where the zero-phonon state (yrast band) is symmetric with respect to $K_{\emph{i}}=0$ and
one-phone state (side band) is antisymmetric~\cite{Qi2009Phys.Lett.B175,Qi2009PhysRevC.79.041302}.

As the angular momentum increases, $K_{\emph{i}}$ for the doublet bands approaches each other at the spin region $14-15\hbar$,
indicating that the chiral vibration through $\emph{s-l}$ is weakening and has a tendency towards the static chirality.
Moreover, the maximum $K_{\emph{s}}$ distribution appears at different position for the doublet bands, shows that the motion contains a vibration of the vector $\emph{I}$ through the $\emph{l-i}$ plane.
Looking at the overall $\emph{K plots}$, with the increasing spin, the peak value of $K_{\emph{l}}\approx4$, while the peak values of both $K_{\emph{i}}$ and $K_{\emph{s}}$ increase slightly and are greater than 4,
indicating that the total angular momentum gradually deviates from the $\emph{l}$ axis.
This is very similar as $^{124}$Cs~\cite{PhysRevC.82.027303}, interpreted as a
typical chiral vibration pattern at the whole spin region based on the present calculations.

\end{multicols}
\vspace{0.5cm}
\begin{center}
\includegraphics[width=16cm]{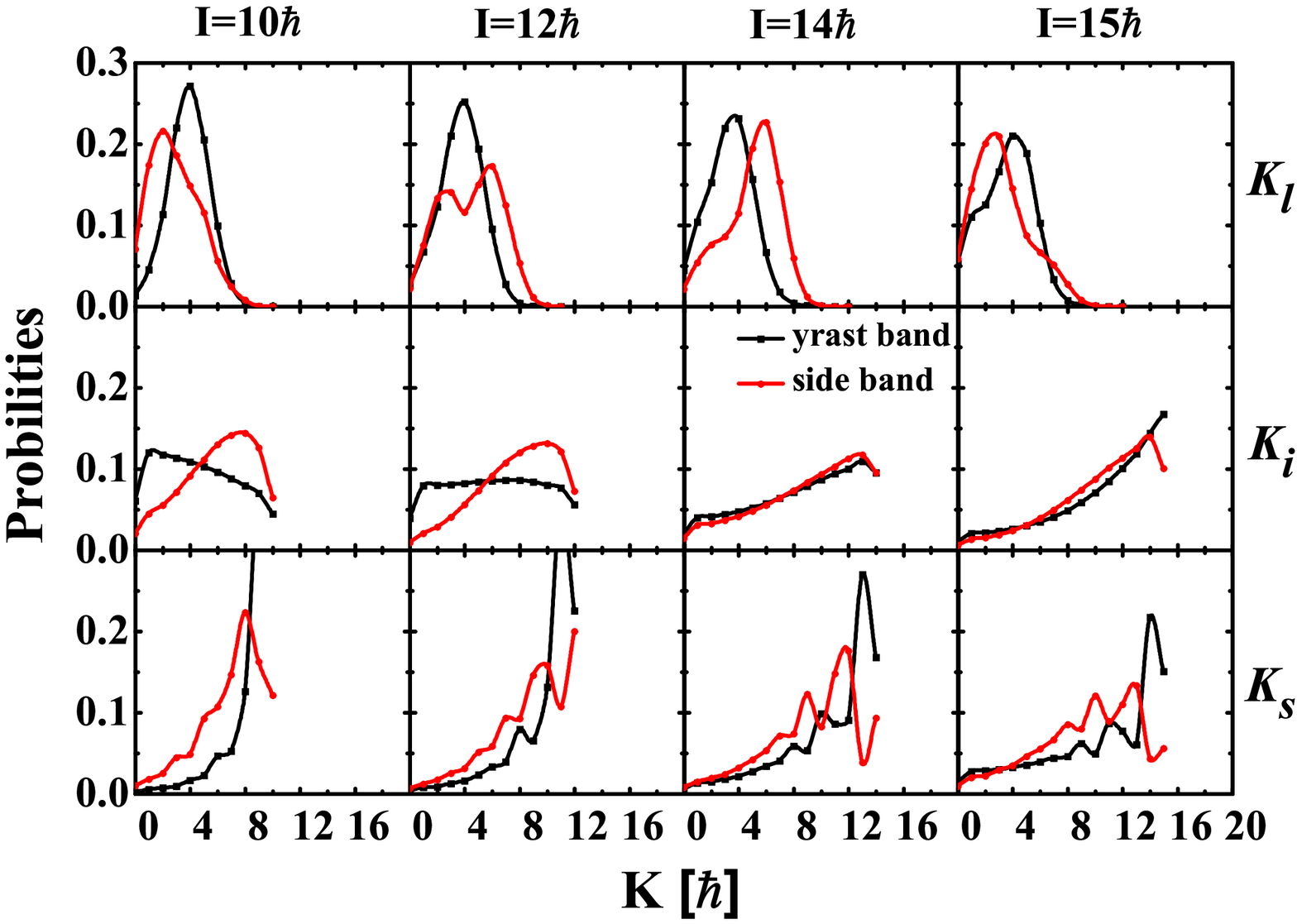}
\figcaption{
(Color online) \label{fig9} The probability distributions for projection $\emph{K}$ of total angular momenta on the long ($\emph{l}$), intermediate ($\emph{i}$), and short ($\emph{s}$) axis in PRM for the doublet bands in $^{120}$I.}
\end{center}
\begin{multicols}{2}

\section{Summary}

Based on the reported positive-parity doublet bands in $^{120}$I, the experimental characteristics including rotational alignment have been extracted and discussed, and the corresponding configuration of the doublet bands is reexamined as $\pi h _{11/2}\otimes\nu h^{-1} _{11/2}$ after performing the TAC-RMF calculations. For the sake of high-$j$ particle hole configuration,
TAC-RMF calculation indicates that $^{120}$I locates at the borders of the $A \approx 130$ island of chiral candidates with configuration $\pi h _{11/2}\otimes \nu h ^{-1}_{11/2}$. However, more neighbouring nuclei should be investigated to obtain more clear conclusion. Moreover, the positive-parity doublet bands based on the $\pi h _{11/2}$ $\otimes$ $\nu h^{-1} _{11/2}$ configuration in $^{120}$I have been studied in two quasiparticles coupled with a triaxial rotor model. The calculated energy spectra, energy staggering parameter $S(I)$, and the intraband $B(M1)/B(E2)$ are in good agreement with the available experimental data. In addition, the calculated effective angles between the angular momenta of the core, valence proton and neutron, and probability distributions for projection of total angular momenta also indicates an obvious chiral geometry of aplanar rotation. Thus we suggest the reported positive-parity doublet bands in $^{120}$I as candidate of chiral doublet bands.

\acknowledgments{
This work is supported by the National
Natural Science Foundation of China under Grant Nos. 11675063, No. 11205068,
No. 11475072, and No. 11847310. The authors would like to thank Prof. J. Meng and S. Q. Zhang
as well as Dr. Q. B. Chen for helpful discussions during the completion of this work.
}

\end{multicols}


\vspace{-1mm}
\centerline{\rule{80mm}{0.1pt}}
\vspace{2mm}

\begin{multicols}{2}

\end{multicols}

\clearpage
\end{CJK*}
\end{document}